\documentclass{aa}

\usepackage{graphicx}
\usepackage{txfonts}

\usepackage{natbib}
\usepackage[colorlinks=true,linkcolor=blue,citecolor=blue]{hyperref}
\bibpunct{(}{)}{;}{a}{}{,} 
	
\title{Determining the time before or after a galaxy merger event}

\author{W.~J.~Pearson\inst{\ref{inst:NCBJ}}
	   \and V.~Rodriguez-Gomez\inst{\ref{inst:IdRA}}
	   \and S.~Kruk\inst{\ref{inst:ESA}}
	   \and B.~Margalef-Bentabol\inst{\ref{inst:SRON}}
}

\institute{National Centre for Nuclear Research, Pasteura 7, 02-093 Warszawa, Poland\label{inst:NCBJ}\\\email{william.pearson@ncbj.gov.pl}
	\and Instituto de Radioastronom\'{i}a y Astrof\'{i}sica, Universidad Nacional Aut\'{o}noma de M\'{e}xico, Apdo. Postal 72-3, 58089 Morelia, Mexico\label{inst:IdRA}
	\and European Space Agency (ESA), European Space Astronomy Centre (ESAC), Camino Bajo del Castillo s/n E-28692 Villanueva de la Ca\~nada, Madrid, Spain\label{inst:ESA}
	\and SRON Netherlands Institute for Space Research, Landleven 12, 9747 AD Groningen, The Netherlands\label{inst:SRON}
}

\date{Received DD Month YYYY; accepted DD Month YYYY}

\abstract{}
{This work aims to reproduce the time before or after a merger event of merging galaxies from the IllustrisTNG cosmological simulation using machine learning.}
{Images of merging galaxies were created in the $u$, $g$, $r$, and $i$ bands from IllustrisTNG. The merger times were determined using the time difference between the last simulation snapshot where the merging galaxies were tracked as two galaxies and the first snapshot where the merging galaxies were tracked as a single galaxy. This time was then further refined using simple gravity simulations. These data were then used to train a residual network (ResNet50), a Swin Transformer (Swin), a convolutional neural network (CNN), and an autoencoder (using a single latent neuron) to reproduce the merger time. The full latent space of the autoencoder was also studied to see if it reproduces the merger time better than the other methods. This was done by reducing the latent space dimensions using Isomap, linear discriminant analysis (LDA), neighbourhood components analysis, sparse random projection, truncated singular value decomposition and uniform manifold approximation and projection.}
{The CNN is the best of all the neural networks. The performance of the autoencoder was close to the CNN, with Swin close behind the autoencoder. ResNet50 performed the worst. The LDA dimensionality reduction performed the best of the six methods used. The exploration of the full latent space produced worse results than the single latent neuron of the autoencoder. For the test data set, we found a median error of 190~Myr, comparable to the time separation between snapshots in IllustrisTNG. Galaxies more than $\approx 625$~Myr before a merger have poorly recovered merger times, as well as galaxies more than $\approx 125$~Myr after a merger event.}
{}


\begin{document}
\authorrunning{W.~J.~Pearson et al.}
\maketitle

\section{Introduction}\label{sec:intro}
Galaxy mergers are a major component of how we understand galaxies to grow and evolve over cosmic time. In our cold dark matter paradigm, dark matter halos merge hierarchically. As a result, the galaxies that they host are also caused to merge \citep[e.g.][]{2014ARA&A..52..291C, 2015ARA&A..53...51S}. This results in short lived morphological changes, such as the creation of tidal tails, as well as longer lived changes, for example changing a galaxy from a late type to early type \citep[e.g.][]{2013ApJ...778...61T}.

These short term morphological changes are exploited to identify galaxy mergers. Visual identification of galaxy mergers relies on astronomers, both professional and amateur, being able to identify distortions and faint structures of galaxies caused by these interactions \citep[e.g.][]{2008MNRAS.389.1179L, 2019AJ....158..103H, 2022A&A...661A..52P}. The disturbances also change the parametric and non-parametric parameters of the merging galaxies, allowing merger selection using concentration, asymmetry, and smoothness \citep[CAS,][]{2000ApJ...529..886C, 2003AJ....126.1183C} or Gini and M$_{20}$ \citep[]{2004AJ....128..163L, 2008ApJ...672..177L, 2015MNRAS.451.4290S, 2019MNRAS.483.4140R}. The close separation and low relative velocities required for galaxies to merge can also be exploited to identify pairs of galaxies that will soon merge \citep[e.g.][]{2000ApJ...530..660B, 2005AJ....130.1516D, 2014MNRAS.444.3986R, 2018MNRAS.475.5133R, 2019ApJ...876..110D}. With morphology based identification typically finding galaxies that have just merged (post-mergers) and the close pair method finding galaxies that are about to merge (pre-mergers), galaxies that are identified as merging by both methods are likely to be currently merging \citep{2023MNRAS.523.4381D}. However, all of these methods suffer from misclassification and contamination in the merging galaxy samples they produce \citep[e.g.][]{2015ApJS..221....8H, 2019A&A...631A..51P}.

To improve merger detection astronomers have turned to machine learning methods. For over half a decade, astronomers have used convolutional neural networks (CNNs) to identify merging and non-merger galaxies in images \citep[e.g.][]{2018MNRAS.479..415A, 2019MNRAS.490.5390B, 2019A&A...631A..51P, 2022A&A...661A..52P, 2019MNRAS.483.2968W, 2020A&C....3200390C, 2020A&A...644A..87W, 2021MNRAS.504..372B, 2022MNRAS.514.3294B}. This methodology also relies on the CNN being able to detect faint structures in the images \citep[e.g.][]{2022A&A...661A..52P}. Machine learning methods have also been applied to morphological parameters \citep{2019MNRAS.486.3702S, 2022A&A...661A..52P, 2023MNRAS.519.4920G, Margalef2024} and photometry \citep{2023A&A...669A.141S} to successfully distinguish merging galaxies from their non-merging counterparts. These methods typically achieve precisions between 75\% and 85\%. More refined classification into pre-mergers, post-mergers and non-mergers is more difficult, with precisions between 65\% and 75\% \citep{2020ApJ...895..115F, Margalef2024}.

Galaxy mergers are also known to influence the physical properties of the merging galaxies. The star-formation rates (SFRs) of galaxies are known to be increased dramatically by interacting galaxies as well as the activity of active galactic nuclei (AGN). However, the exact change seen in these properties is contested. Mergers have been seen to trigger extremely enhanced SFR (star-bursts) in some studies \citep[e.g.][]{1996ARA&A..34..749S, 2019A&A...631A..51P}, while other studies have found the enhancement to be too low to be considered a star-burst, with an increase of at most a factor of two \citep[e.g.][]{2013MNRAS.435.3627E, 2018ApJ...868...46S, 2019A&A...631A..51P}, or indeed see a reduction in SFR \citep{2015MNRAS.454.1742K}. Similarly, the increase in AGN activity is found in some studies \citep[e.g.][]{2017MNRAS.464.3882W, 2020A&A...637A..94G, 2023ApJ...942..107S} and decrease in others \citep[e.g.][]{2021ApJ...909..124S}.

These differences in changes of physical properties are likely a result of the stage of merger that is identified. The FIRE-2 zoom-in simulations \citep{2018MNRAS.480..800H} have found that the SFR of merging galaxies is dependent on the time before and after a merger event \citep{2019MNRAS.485.1320M}. Similarly, AGN activity is found to decrease the more time has elapsed since a merger event in the IllustrisTNG cosmological simulation \citep{2018MNRAS.475..648P, 2023MNRAS.519.4966B}. However, beyond pre-merger and post-merger separation, it is currently not simple to determine the time before or after a merger event. With pre-merger, post-merger, and non-merger classification harder than merger non-merger classification, we expect merger time estimation to be more difficult still.

Determining the merger time can be done using machine learning and simulations. Simulations provide us with the time a galaxy is before or after a merger, although not without limitations. Zoom-in simulations of merging galaxies provide high time resolution but due to their expense do not provide large samples of galaxies. Cosmological simulations can provide tens or hundreds of thousands of galaxy mergers but have poor time resolution, of order of 100~Myr.

Studies into determining the merger time have barley begun. \citet{2021arXiv210205182K} were able to use the Horizon-AGN cosmological simulation \citep{2014MNRAS.444.1453D} to train a neural network to predict the time before or after a merger event for simulated Hubble Space Telescope observations of mergers. While methods to identify galaxy mergers have started to converge in their quality, more study is required to refine merger time estimation.

In this paper, we aim to compare several methods that determine the time before or after a merger event a galaxy is. We will use merging galaxies identified in IllustrisTNG and perform simple gravity simulations to improve their time resolution. We will use the images of the galaxies and their merger times to train four different neural network architectures and compare their results.

The paper is structured as follows. Section \ref{sec:data} describes our data, Sect. \ref{sec:dl} discusses the deep learning methods used, Sect. \ref{sec:res} presents our results, while Sect. \ref{sec:discuss} discusses our findings. We conclude in Sect. \ref{sec:conc}. Where necessary, we follow the Planck2015 cosmology \citep{2016A&A...594A..13P}.

\section{Data}\label{sec:data}
\subsection{IllustrisTNG}\label{sec:data:tng}
The time before or after a merger event (merger time) is inherently unknown for galaxies in the Universe. However, these merger times are easier to know for galaxies within cosmological simulations, albeit with certain limitations.

Here, we use galaxy mergers identified in the IllustrisTNG simulation's TNG100-1 \citep{2018MNRAS.480.5113M, 2018MNRAS.477.1206N, 2018MNRAS.475..624N, 2019ComAC...6....2N, 2018MNRAS.475..648P, 2018MNRAS.475..676S}. TNG100-1 simulates 1820$^{3}$ dark matter particles within a box of 75~cMpc/h per side and softening length of 0.5~ckpc/h using Planck2015 cosmology. Within this box, the gas cells have an average mass of $9.44 \times 10^{5}$ M$_{\odot}$/h while the dark matter particles have a mass of $50.56 \times 10^{5}$ M$_{\odot}$/h. TNG100-1 has the highest resolution of the TNG100 simulations. The times between simulation snapshots are from 30~Myr to 234~Myr with an average separation of 138~Myr. Larger times between snapshots are typically found at lower redshifts. The galaxies from IllustrisTNG are also known to be morphologically similar to observed galaxies, with trends of morphology, size and shape with stellar mass within the 1$\sigma$ scatter of observational trends \citep{2019MNRAS.483.4140R}. However, the morphology-colour relation is not strong in IllustrisTNG and disk galaxies are not found to be larger than ellipticals at fixed stellar mass, unlike galaxies observed in the real universe \citep{2019MNRAS.483.4140R}.

We define a merging galaxy as a galaxy that has merged in the last 500~Myr (post-merger) or will merge within the next 1000~Myr (pre-merger), as determined by the snapshot of merger relative to the snapshot in which the galaxy is identified \citep[snapshot of observation;][]{2020A&A...644A..87W} using the SubLink\_gal merger trees \citep{2005Natur.435..629S, 2009MNRAS.398.1150B, 2015MNRAS.449...49R}. The snapshot of merger is the snapshot in which a single galaxy was tracked as two, or more, galaxies in the previous snapshot. We define a major merger as a merger where the ratio between the stellar masses of the most massive and second most massive galaxies is $< 1:4$ in the snapshot where the second most massive galaxy reached its maximum stellar mass \citep{2015MNRAS.449...49R}. {We require the mass of the baryonic component (stars + star-forming gas) of the merging galaxy's subhalo to be at least $1\times10^{9}$~M$_{\odot}$/h. Major merging galaxies that meet both these criteria are then selected from snapshots 87 to 93 (inclusive), corresponding to $0.07 \leq z \leq 0.15$. This range is chosen as the upper redshift limit corresponds to the upper limit of the \citet{2019A&A...631A..51P} galaxy merger catalogue in the Kilo Degree Survey \citep[KiDS;][]{2013Msngr.154...44D, 2013ExA....35...25D}. The lower redshift limit allows for our full pre-merger time period before redshift 0 and therefore provides a more complete pre-merger sample. If a galaxy is found to be merging in multiple snapshots, the galaxy will be selected in each snapshot it is found to be merging in within our interval and so may appear multiple times. This resulted in 6139 major mergers being selected.

To increase the time-resolution of the merger times, simple gravity simulations are run to determine when the galaxies merged. For this, each galaxy was considered a point mass and was allowed to move through space only under the gravitational influence of other galaxies involved, using a softening length of 0.5~ckpc/h to match TNG100. The merger was simulated using all galaxies with a mass ratio of $< 1:10$ to the primary galaxy. Including less massive galaxies in the simulation allows the smaller galaxies to influence the interaction. The simulation was run with a time step of 1~kyr over the time between snapshots. The merger was deemed to have occurred when the major merger galaxies reached their first closest approach. If there were more than two major galaxies in a merger, the first closest approach of the last galaxy to closely approach the primary galaxy was used as the merger moment for post mergers while the first closest approach of any major galaxy was used as the merger moment for the pre-mergers. The resulting merger time was then rounded to the closest 1~Myr. If the gravity simulation did not provide a closest approach, the galaxy was removed from further study. We also remove any galaxy that is classified as both a pre-merger and post-merger, to exclude potentially ambiguous cases. This resulted in the removal of 2818 galaxies, leaving 3321 galaxies with a merger time with redshifts from 0.07 to 0.15.

We define merger time as ``time before a merger'', therefore if a galaxy is a post-merger, its merger time will be negative. The merger time is scaled between 0 and 1, such that a merger time of -500~Myr is 0 and 1000~Myr is 1. These scaled merger times are referred to as normalised merger times here on. As can be seen in Fig. \ref{fig:data:time-separation}, the normalised merger time for pre-mergers is not strongly correlated with the projected separation between the merging galaxies at the time of the simulation when the images were generated (Pearson coefficient = 0.277, p-value = 0.000; Spearman correlation = 0.407, p-value = 0.000). We generate three images of each galaxy, viewed along the x, y, and z axes of the simulation, giving 9963 images in total.

\begin{figure}
	\resizebox{\hsize}{!}{\includegraphics{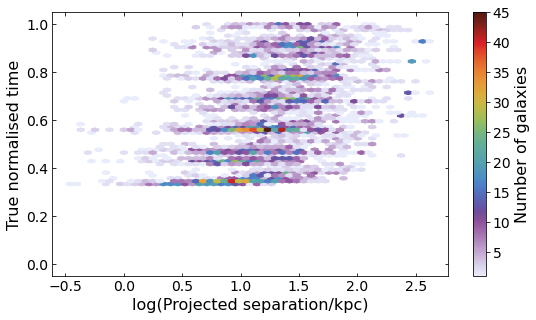}}
	\caption{True merger times against projected separations of merging galaxies at the time of imaging for pre-mergers. Colour corresponds to number density from low (purple) to high (red).}
	\label{fig:data:time-separation}
\end{figure}

To create the images of the merging galaxies, each stellar particle in the simulation has a spectral energy distribution derived from the \citet{2003MNRAS.344.1000B} stellar population models and the stellar properties of the particle. The resulting spectra are passed through the desired filters, here KiDS $u$, $g$, $r$, and $i$ bands, to create a smoothed 2D image \citep{2019MNRAS.483.4140R}. The pixel resolution of the IllustrisTNG images is that of the KiDS images, 0.2~arcsec/pixel, and we produce cut-outs of 128$\times$128 pixels centred on the IllustrisTNG galaxies. We do not enforce that the companion galaxy of a pre-merger is inside the image as we expect the disturbances of pre-merger and post-merger galaxies to be different and hence distinguishable. It has been shown the extra realism from the calculation of the reprocessing the light due to dust does not notably increase the performance of a neural network in identifying galaxy mergers \citep{2019MNRAS.490.5390B}. It also does not cause large changes to the morphology of a galaxy \citep{2019MNRAS.483.4140R}. Thus, radiative transfer was not calculated for these galaxies. Examples of the images of the merging galaxies are shown in Fig. \ref{fig:data:tng}.

\begin{figure}
	\resizebox{\hsize}{!}{\includegraphics{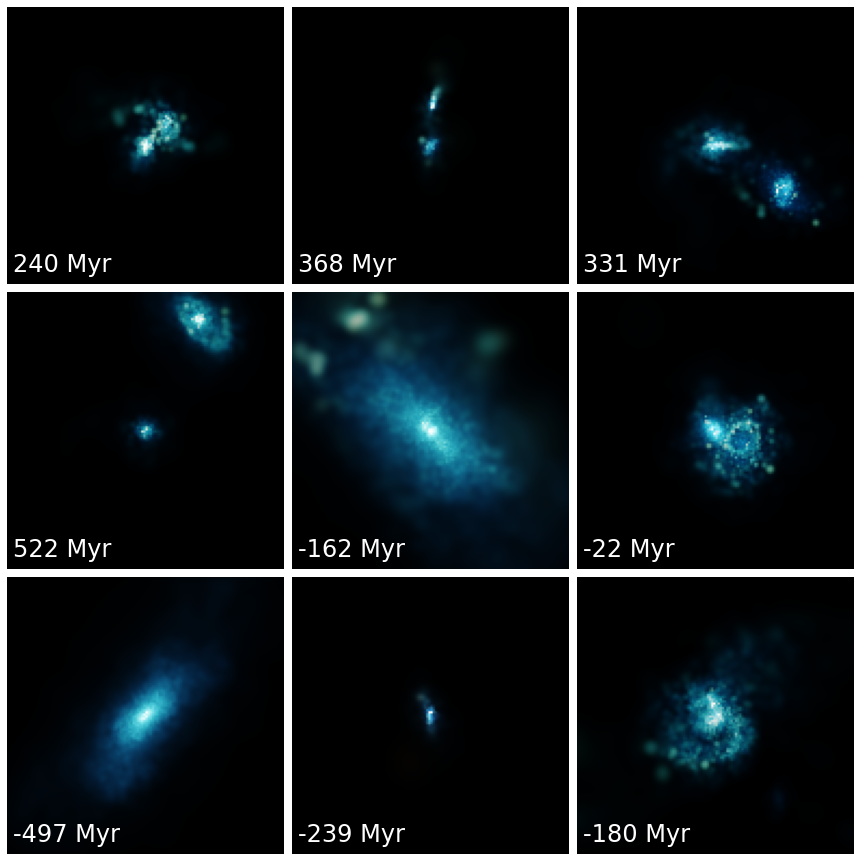}}
	\caption{Example $u$, $g$, and $r$ band composite images of merging IllustrisTNG galaxies used in this work. Each channel is individually arcsinh scaled twice and then normalised between 0 and 1. Merger times of each galaxy are shown in the bottom left of each panel.}
	\label{fig:data:tng}
\end{figure}

\section{Neural networks}\label{sec:dl}
Neural networks are a subset of machine learning that aims to loosely replicate how biological, neurological systems process information. In this work, we perform supervised regression; that is we train a network using data with known truth values to generate a value from a continuous distribution. To do this we employ a Residual Network \citep{2015arXiv151203385H}, a Swin Transformer \citep{2021arXiv211114725C, liu2021Swin}, a convolutional neural network (CNN), and an autoencoder. The data used to train a network in supervised learning are typically subdivided into three subsets: a training set that typically contains 70-90\% of all the data that is used to train the network; a validation set that typically contains 5-15\% of all the data that is used to check the performance of the network as it trains; and a test set that typically contains 5-15\% of all the data that is used once, and only once, to check the performance of the network once training is complete. The exact split between these three subsets is a matter of choice and can be tuned to help training; in this work we use a split of 80\%-10\%-10\% for the training, validation and test data, respectively.

For each architecture tested, we augment the images during training, but not validation. This is done by randomly rotating the image by a multiple of 90$^{o}$. This rotation is then followed by randomly flipping the image horizontally and randomly flipping the image vertically. Doing so increases the effective size of the training set by a factor of 16, although the augmented images are not as informative as entirely new data. Each band of each image is also individually arcsinh scaled twice then normalised between 0 and 1. As the ultimate aim is to apply these techniques to real observations, we normalise the bands separately to remove connections between bands. Simulations are not a perfect replication of the real universe so training with connected bands may result in the networks learning connections between bands that are not present in observations.

The different architectures were compared using mean squared error (MSE) calculated between the true merger time and the predicted merger time. As will be seen, not all architectures were trained using MSE. All architectures are built within the Tensorflow machine learning framework \citep{tensorflow2015-whitepaper}. We train each architecture for 1000 epochs and select the epoch with the lowest MSE for the merger times with the validation set. The codes to create the architectures along with the trained models, can be downloaded from GitHub\footnote{\url{https://github.com/wjpearson/TNGMergerTime}}.

\subsection{ResNet50}\label{sec:dl:resnet}
Residual networks \citep[ResNet,][]{2015arXiv151203385H} use residual learning to allow deep networks to train with degradation similar to a shallower network. This is achieved by introducing skip connections between convolutional blocks, allowing the input to both pass through the convolutional block as well as skip it entirely. These architectures were originally conceived for image classification but have been used previously for regression in astronomy \citep[e.g.][]{2021arXiv210205182K}. Here we use a ResNet with 50 convolutional blocks (ResNet50), we remove the fully connected top layers, and add a single output neuron with sigmoid activation. The input is a three channel $128\times128$ pixel image using the $u$, $g$, and $r$ bands. The choice of bands is arbitrary as it has no impact on the results. The network is trained with MSE loss using the Adam optimiser \citep{2014arXiv1412.6980K}.

\subsection{Swin Transformer}\label{sec:dl:swin}
We use a Swin Transformer (Swin) architecture \citep{2021arXiv211114725C, liu2021Swin} that has been pre-trained on ImageNet-1K \citep{ImageNet1k} using a publicly available model\footnote{\url{https://github.com/sayakpaul/swin-transformers-tf}}. We add a single output neuron to the Swin Transformer to provide classification. As we are using a pre-trained network, we require the input images to have three channels and be $224\times224$ pixels. This was achieved by using the $u$, $g$, and $r$ bands to create a three channel image and cropping the image to $112\times112$ pixels about the centre. Again choice of bands has no impact on the results. These cropped images were then resized to $224\times224$ pixels using nearest neighbour interpolation. The output of the network was a value between 0 and 1 using sigmoid activation, being a scaled version of the merger time. The network was trained with MSE loss using a stochastic gradient descent (SGD) optimiser \citep{10.1214/aoms/1177729586, 10.1214/aoms/1177729392} with a warm up cosine\footnote{\url{https://www.kaggle.com/ashusma/training-rfcx-tensorflow-tpu-effnet-b2}}.

\subsection{CNN}\label{sec:dl:cnn}
For our CNN, we use six convolutional layers followed by three fully connected layers and a single neuron output, as shown in Fig. \ref{fig:dl:cnn}. The convolutional layers have 32, 64, 128, 256, 512, and 1024 filters with a size of 6, 5, 3, 3, 2, and 2 pixels, respectively, and Rectified Linear Units \citep[ReLU,][]{Nair2010} activation. Each convolutional layer uses a stride of 1 and ``same'' padding. We follow each convolutional layer with batch normalisation, dropout with a rate of 0.2 for convolutional layers and 0.1 for fully connected layers, and $2\times2$ max-pooling. The fully connected layers have 2048, 512, and 128 neurons and ReLU activation. Like the convolutional layers, they are followed by batch normalisation and dropout with a rate of 0.1. The input is a four channel ($u$, $g$, $r$, and $i$ bands), $128\time128$ pixel image while the output later is a single neuron with sigmoid activation. The CNN was trained using MSE loss and the Adam optimiser.

\begin{figure*}
	\resizebox{\hsize}{!}{\includegraphics{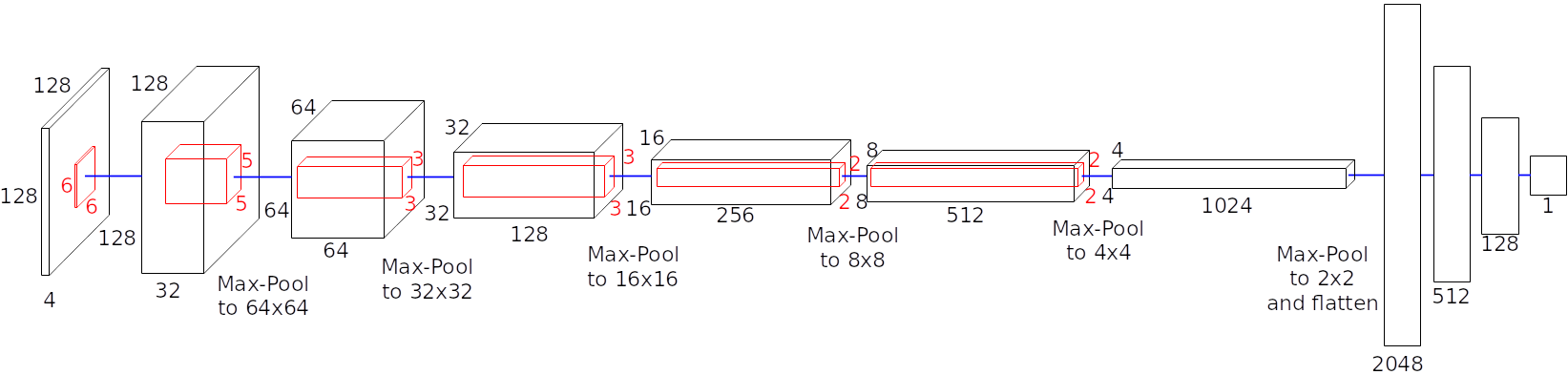}}
	\caption{Architecture of the CNN. The input to the CNN is a four channel, $128\times128$ pixel image, on the left of this schematic. The output is a single neuron, on the right of the schematic. The blue lines between layers symbolises the ReLU activation, batch normalisation, and dropout that is applied between layers. The sizes of the filters (red) and fully connected layers are shown.}
	\label{fig:dl:cnn}
\end{figure*}

\subsection{Autoencoder}\label{sec:dl:autoencoder}
The autoencoder uses the CNN architecture to encode the images into a latent space with 64 neurons with sigmoid activation, which replaces the single neuron output of the CNN, as seen in Fig. \ref{fig:dl:autoencoder}. The decoder begins with four fully connected layers of 128, 512, 2048, and 4096 neurons with ReLU activation. Each fully connected layer is followed by batch normalisation and dropout with a dropout rate of 0.1. The output of the 4096 layer is reshaped to $2\times2\times1024$. This is followed by six transposed convolutions with 1024, 512, 256, 128, 64, and 32 filters with sizes of 2, 2, 3, 3, 5, and 6 pixels. Before the transposed convolutions there is a $2\times2$ upsampling layer and the transposed convolutions are followed by batch normalisation and dropout with a dropout rate of 0.2. The final layer is a transposed convolution with 4 filters of size $1\times1$ pixels and sigmoid activation.

\begin{figure*}
	\resizebox{\hsize}{!}{\includegraphics{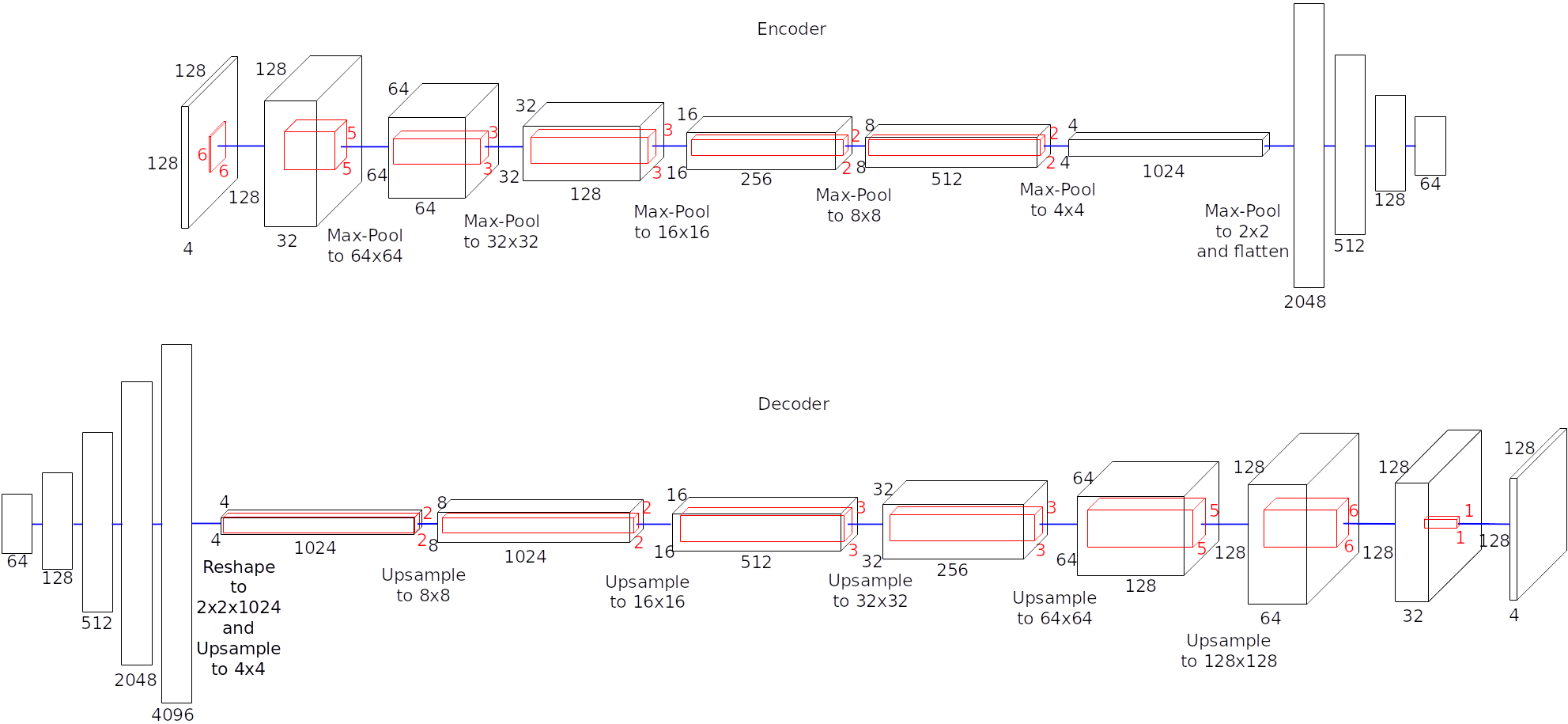}}
	\caption{Architecture of the autoencoder. For clarity, the encoder is shown at the top and the decoder at the bottom. In reality the 64 neuron output of the encoder, on the upper right, and 64 value input to the decoder, on the lower left, are the same layer. The input to the encoder is a four channel, $128\times128$ pixel image, on the left of the encoder. The output from the decoder, on the right, is a four channel, $128\times128$ pixel image. The blue lines between layers symbolises the ReLU activation, batch normalisation, and dropout that is applied between layers. The sizes of the filters (red) and fully connected layers are shown.}
	\label{fig:dl:autoencoder}
\end{figure*}

Here, the goal is not to perfectly reproduce the images after decoding. Instead, we aimed to use the latent space to hold information about the merger time. This was done by training one of the latent neurons to reproduce the merger time. We were running into issues with the CNN where all output values tended to 0.5, and it was hoped that allowing data to flow around the output neurone, similar to a ResNet, may prevent this. The latent space encoding also allows us to investigate if the entire latent space holds more information about the merger time than the single output neurone. Like the CNN, the Adam optimiser was used while the loss was the sum of the MSE of the input and recreated images summed with the MSE of the merger time latent neuron.

\subsection{Latent space}\label{sec:dl:latent}
We investigate the latent space of the autoencoder to determine if there is information that can be used to determine the time before or after a merger event a galaxy merger is. We do this using a number of different dimensionality reduction techniques, to reduce the 64-dimension latent space into a 2-dimensional representation. We require the dimensionality reduction techniques to allow new data to be placed into the same dimension reduction mapping. This allows the training data to be used to create the mapping and the validation data to validate the methodology. We use the \texttt{Scikit-learn} \citep{scikit-learn} implementation of the majority of the following dimensionality reduction methods. For all methods that we use the \texttt{Scikit-learn} implementation of, we use the default configuration but set \texttt{n\_components} to 2. For Uniform Manifold Approximation and Projection \citep[UMAP,][]{2018arXiv180203426M}, we use the \texttt{UMAP-learn} implementation\footnote{\url{https://umap-learn.readthedocs.io/en/latest/index.html}} with the default configuration.

\begin{itemize}
	\item Isomap \citep{2000Sci...290.2319T}
	\item Linear discriminant analysis (LDA). LDA is a generalised form of Fisher's linear discriminant \citep{fisher1936use}.
	\item Neighbourhood Components Analysis \citep[NCA,][]{NIPS2004_42fe8808}.
	\item Sparse Random Projection \citep[SRP,][]{10.1145/1150402.1150436}
	\item Truncated singular value decomposition \citep[TSVD,][]{doi:10.1137/090771806}.
	\item UMAP
\end{itemize}

To determine the times of galaxies without known merger times, we use the latent space mapping to embed the unknown data into the mapping of the training data. We identify the three closest training points to a unknown point that form a triangle around the unknown point in the embedded space. Using the known merger time of the training data as a third axis, we form a 3-dimensional plane with these three training data. The time for the unknown datum is then the z-position on the plane for the unknown datum's embedding. The codes to perform the latent space embedding along with the embeddings can be downloaded from GitHub\footnote{\url{https://github.com/wjpearson/TNGMergerTime}}.

\section{Results}\label{sec:res}
\subsection{Neural networks}\label{sec:res:nn}
We present the loss of the neural networks in Table \ref{tab:res:dl} and as blue symbols in Fig. \ref{fig:summary}. As can be seen, the CNN architecture provides the lowest MSE of the merger times and hence is the best performing architecture. The CNN is of course only slightly better performing than the autoencoder, and retraining both architectures may result in the autoencoder performing better than the CNN. The predicted versus true normalised times for all four architectures can be found in Fig. \ref{fig:res:time-nn}

\begin{table}
	\centering
	\caption{MSE of the normalised merger times for different architectures from the epoch with the lowest MSE of the normalised merger times of the validation data set, along with their associated mean and median un-normalised merger time errors.}
	\label{tab:res:dl}
	\begin{tabular}{ccccc}
		\hline
		\hline
		Architecture & Training & Validation & \multicolumn{2}{c}{Validation Error}\\
		 & MSE & MSE & Mean\tablefootmark{a} & Median\tablefootmark{a} \\
		\hline
		ResNet50 & 0.075 & 0.065 & 327 & 324\\
		Swin & 0.034 & 0.042 & 236 & 193\\
		CNN & 0.040 & 0.037 & 215 & 165 \\
		Autoencoder & 0.036 & 0.038 & 222 & 160\\
		\hline
	\end{tabular}
	\tablefoot{
		\tablefoottext{a}{Values in Myr}
	}
\end{table}

\begin{figure}
	\resizebox{\hsize}{!}{\includegraphics{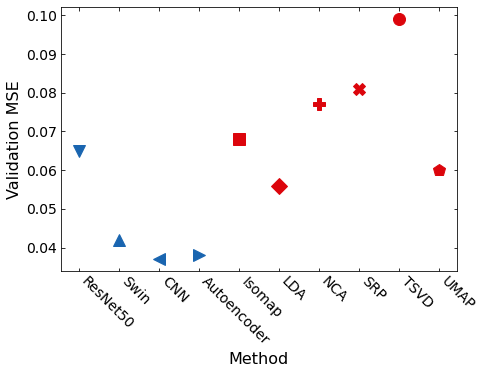}}
	\caption{MSE at validation for the neural networks (blue symbols) ResNet50 (downward pointing triangle), Swin (upward pointing triangle), CNN (left pointing triangle), and autoencoder (right pointing triangle) as well as the latent space embeddings (red symbols) Isomap (square), LDA (diamond), NCA (plus), SRP (cross), TSVD (circle), and UMAP (pentagon).}
	\label{fig:summary}
\end{figure}

\begin{figure}
	\resizebox{\hsize}{!}{\includegraphics{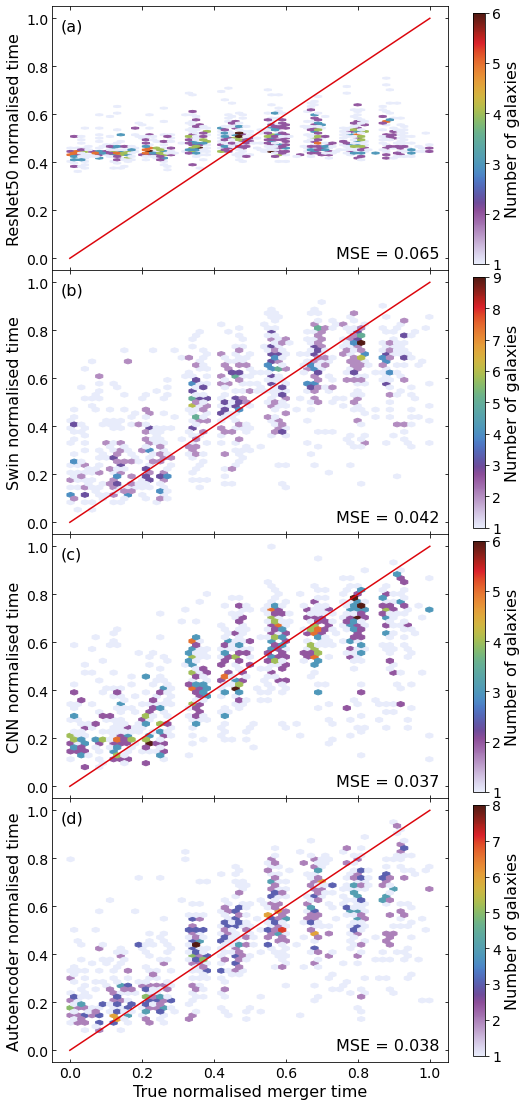}}
	\caption{Predicted normalised times of the validation data against true normalised times for (a) ResNet50, (b) Swin, (c) CNN, and (d) autoencoder. Colour corresponds to number density from low (purple) to high (red). Red line indicates 1-to-1 and corresponding MSE is shown in the bottom right of each panel.}
	\label{fig:res:time-nn}
\end{figure}

We also examine how the mean and median absolute difference between the true times and predicted time (error) varies at different true time for each network in Fig. \ref{fig:res:time-error}. For this, we bin the normalised true times into bins of width 0.1 and calculate the mean and median errors within each bin. The mean and median errors closely follow the same trend for each network. As can be seen, the ResNet50 mean and median decrease towards a normalised time of $\approx$0.5 before increasing again; a result of the network only outputting values close to 0.5. The other neural networks have sharp decreases in errors at the low end of the normalised merger time range ($\lesssim 0.25$) and sharp increases in errors at the high end ($\gtrsim 0.75$). These other networks then have better performance between these normalised true times of 0.25 and 0.75, where the curves are flatter and lower in values. We also see a slight increase in errors around normalised merger times of 0.3, where the merger event happens.

\begin{figure}
	\resizebox{\hsize}{!}{\includegraphics{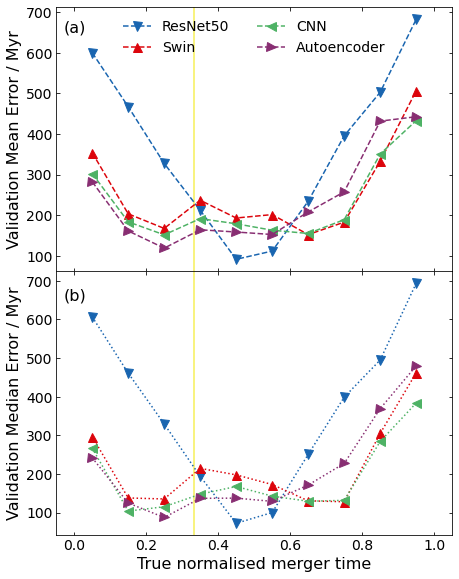}}
	\caption{(a) Mean and (b) median difference between the true merger time and predicted merger time (error) as the normalised true merger time changes for ResNet50 (dark blue downward pointing triangle), Swin (red upward pointing triangle), CNN (green left pointing triangle), and autoencoder (purple right pointing triangle). Vertical yellow line indicates when the true merger time is 0~Myr.}
	\label{fig:res:time-error}
\end{figure}

\subsection{Latent space}\label{sec:res:latent}
The latent space embedding for LDA can be found in Fig. \ref{fig:res:lda}, while the embeddings for Isomap, NCA, SRP, TSVD, and UMAP can be found in Appendix \ref{app:embeddings}, using the latent space from the autoencoder epoch with the lowest merger time MSE. This is for both the training data (Fig. \ref{fig:res:lda}a) used to create the embedding and the validation data (Fig. \ref{fig:res:lda}b). In all embeddings, the merger times of the validation data appear to correspond with the merger times of the training data.

\begin{figure}
	\resizebox{\hsize}{!}{\includegraphics{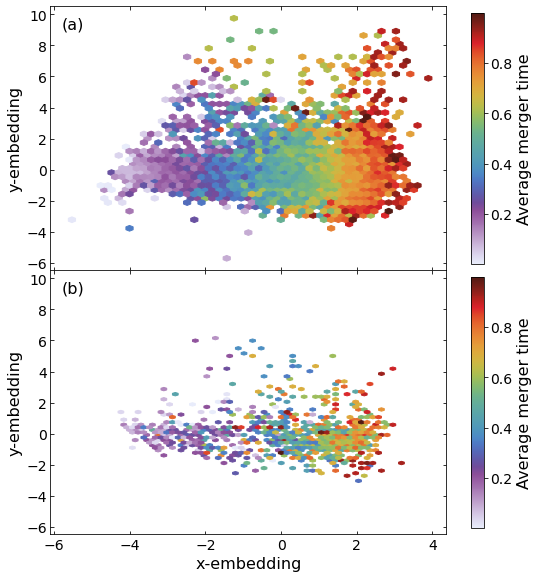}}
	\caption{LDA latent space embedding for (a) training data and (b) validation data. The average normalised merger time is show from 0 (purple) to 1 (red).}
	\label{fig:res:lda}
\end{figure}

Using the method described in Sect. \ref{sec:dl:latent}, we estimate the merger time from each embedding for the validation data. The MSE of these results are shown in Table \ref{tab:res:latent}, and red symbols in Fig. \ref{fig:summary}, and the predicted versus true normalised times can be found in Fig. \ref{fig:res:time-latent}. As can be seen, none of the embeddings produce a MSE lower than the autoencoder and only LDA and UMAP perform better than the worse performing architecture, ResNet50.

\begin{table}
	\centering
	\caption{MSE, mean error, and median error of the merger times of the validation data set for the different latent space embeddings.}
	\label{tab:res:latent}
	\begin{tabular}{cccc}
		\hline
		\hline
		Embedding & Validation & \multicolumn{2}{c}{Validation Error} \\
		 & MSE & Mean\tablefootmark{a} & Median\tablefootmark{a}\\
		\hline
		Isomap & 0.068 & 291 & 231\\
		LDA & 0.056 & 283 & 226\\
		NCA & 0.077 &  332 & 277\\
		SRP & 0.081 & 338 & 287\\
		TSVD & 0.099 & 378 & 318\\
		UMAP & 0.060 & 282 & 223 \\
		\hline
	\end{tabular}
	\tablefoot{
		\tablefoottext{a}{Values in Myr}
	}
\end{table}

\begin{figure*}
	\resizebox{\hsize}{!}{\includegraphics{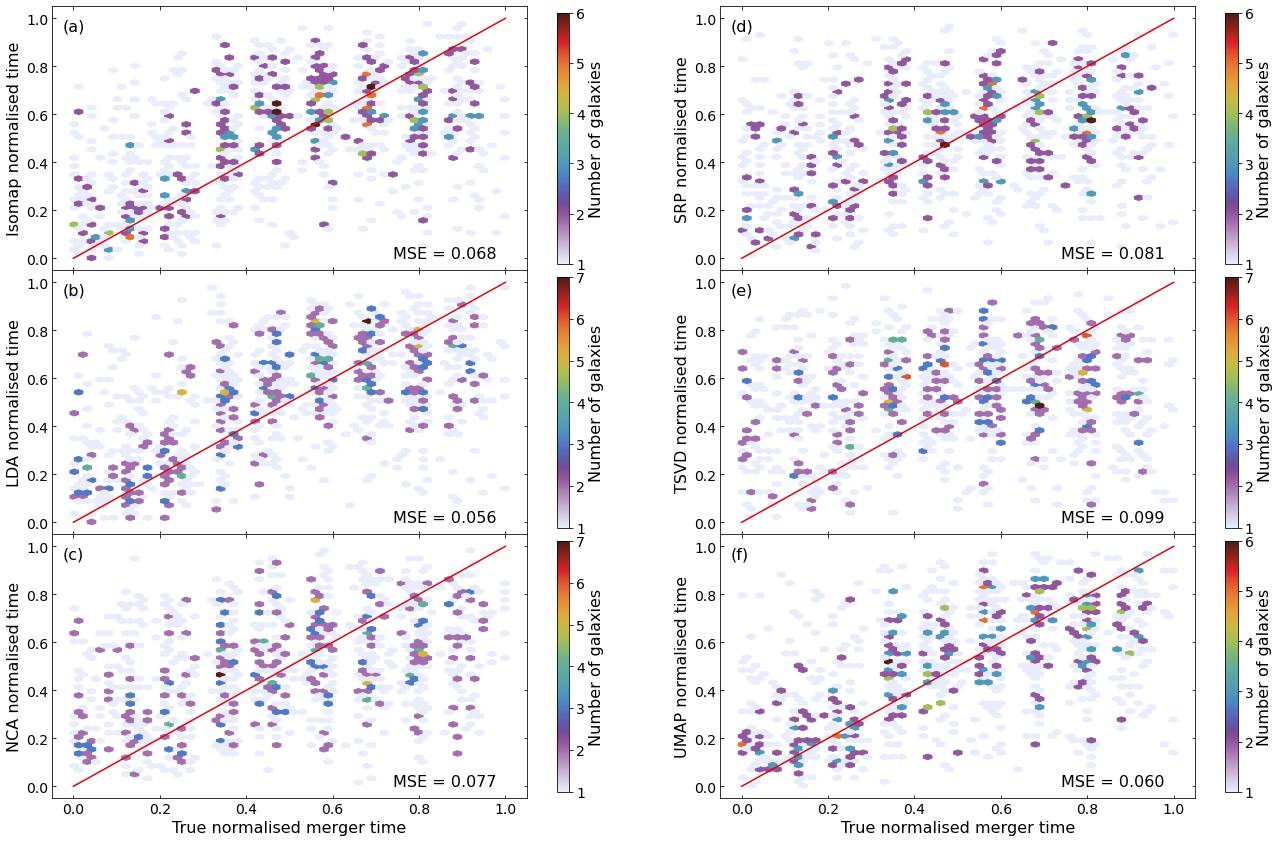}}
	\caption{Predicted normalised merger times of the validation data against true normalised times for (a) Isomap, (b) LDA, (c) NCA, (d) SRP, (e) TSVD, and (f) UMAP. Colour corresponds to number density from low (purple) to high (red). Red line indicates 1-to-1 and the corresponding MSE is shown in the bottom right of each panel.}
	\label{fig:res:time-latent}
\end{figure*}

As with the neural networks, we also study how the mean and median error varies at different true time for each embedding in Fig. \ref{fig:res:latent-error}. The trends broadly fit into two groups. Isomap, LDA, and UMAP have mean errors that decrease to 0.25, then increase to 0.35, decrease again to 0.65, before increasing to 1.0. The median error for these three embeddings decreases to 0.15, increases to between 0.35 and 0.45, decreases to 0.65, before again increasing to 1.0. The other group, NCA, SRP, and TSVD, have a decreasing mean and median error to 0.55 and increasing error from then on.

\begin{figure}
	\resizebox{\hsize}{!}{\includegraphics{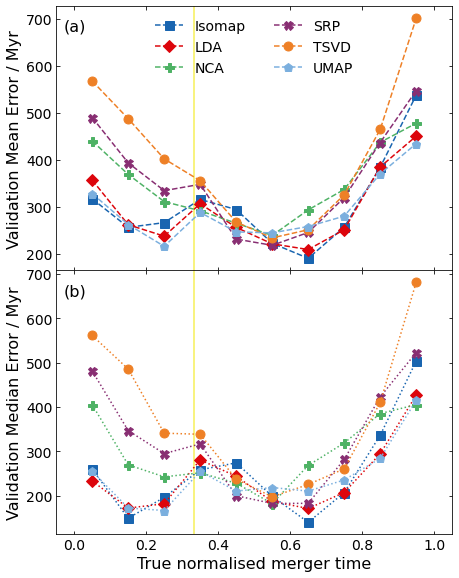}}
	\caption{(a) Mean and (b) median difference between the true merger time of the validation data and predicted merger time (error) as the normalised true merger time changes for Isomap (dark blue squares), LDA (red diamonds), NCA (green pluses), SRP (purple crosses), TSVD (orange circles), and UMAP (light blue pentagons). Vertical yellow line indicates when the true merger time is 0~Myr.}
	\label{fig:res:latent-error}
\end{figure}

As can be seen in Fig. \ref{fig:res:lda}, the LDA x-embedding appears to strongly correlate with the merger time. Thus, we map the LDA x-embedding to the merger time using a linear fit to the training data and calculate the MSE. This gives a training MSE of 0.032 and a validation MSE of 0.039, the latter of which is shown in Fig. \ref{fig:res:time-ldax} as the predicted merger time as a function of true merger time. These results are only slightly worse than the autoencoder validation MSE.

\begin{figure}
	\resizebox{\hsize}{!}{\includegraphics{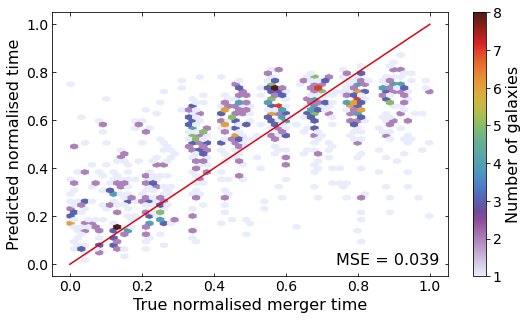}}
	\caption{Predicted normalised merger times of the validation data against true normalised times for LDA x-embedding. Colour corresponds to number density from low (purple) to high (red). Red line indicates 1-to-1 and the corresponding MSE is shown in the bottom right of each panel.}
	\label{fig:res:time-ldax}
\end{figure}

\subsection{Test data}\label{sec:res:test}
With the CNN performing the best of all architectures and latent space explorations, we apply this network to the test data as shown in Fig. \ref{fig:res:test-time}. This results in a MSE of 0.051, an average error of 253~Myr, and a median error of 190~Myr. We present the mean and median errors as a function of true normalised merger time for the test data in Fig. \ref{fig:res:test-error}. These mean errors closely follow the CNN validation mean errors between normalised times of 0.25 and 0.55 with means outside of this range being higher than those found with the validation data. The median test errors are again typically larger than the validation errors, with the exception of the two bins between normalised true times of 0.2 and 0.4. Taking pre-merger to be galaxies with true normalised times greater than 0.33 and post-mergers with true normalised times less than 0.33, we find 81$\pm$1\% of our galaxies are correctly identified, with errors from bootstrapping.

\begin{figure}
	\resizebox{\hsize}{!}{\includegraphics{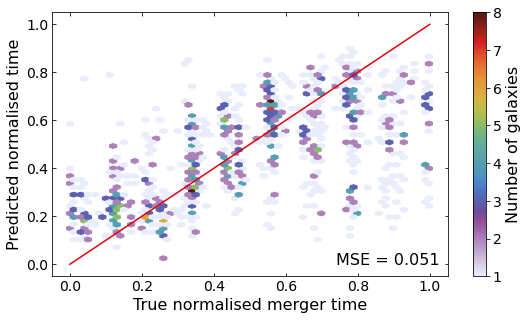}}
	\caption{Predicted normalised merger times from the CNN against true normalised times for the test data. Colour corresponds to number density from low (purple) to high (red). Red line indicates 1-to-1 and the corresponding MSE is shown in the bottom right.}
	\label{fig:res:test-time}
\end{figure}

\begin{figure}
	\resizebox{\hsize}{!}{\includegraphics{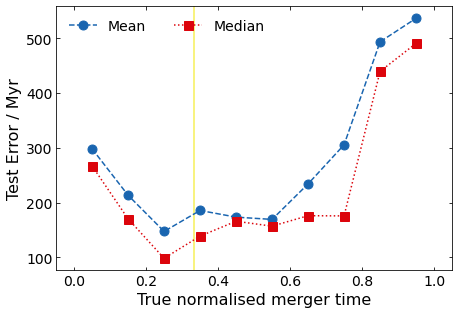}}
	\caption{Mean (dark blue circles) and median (red squares) difference between the true and predicted merger times (error) as the normalised true merger time changes the test data processed by the CNN. Vertical yellow line indicates when the true merger time is 0~Myr.}
	\label{fig:res:test-error}
\end{figure}

\section{Discussion}\label{sec:discuss}
\subsection{Neural networks}\label{sec:discuss:nn}
ResNet50 had the notably worst performance of all architectures tested and only produces values close to 0.5. This is not due to over-fitting, as the training data also exhibits the same problem. This is surprising due to previous success with ResNet50 at determining merger time \citep{2021arXiv210205182K}. However, the work of \citet{2021arXiv210205182K} use over 200\,000 images of merging systems, compared to our 9963 examples, which may be providing the better results. They also use images from the Horizon-AGN cosmological simulation \citep{2014MNRAS.444.1453D}, although we do not expect this to be the cause of our much poorer results. \citet{2021arXiv210205182K} also include the mass and redshift of the merging galaxies, which may provide further information to aid merger time estimation.

Swin has been shown to be useful in identifying galaxy mergers, as well as identifying pre- and post-merging galaxies. This architecture has also been shown to perform favourably in these tasks when compared to a three different CNN structures \citep{Margalef2024}. However, it performs slightly worse than the CNN structure used in this work. Therefore, while this architecture is able to extract information from the images, it does not extract as much as the CNN and autoencoder.

This slight lack of performance may be due to the fewer bands used to create the images used to train Swin. Due to the limitations of using a pre-trained network, we were required to use only three of the four available bands. On the other hand, as the CNN and autoencoder were not pre-trained, and were custom built for this task, we were able to use all four bands we had available. The fourth band may therefore carry further information about the merger time that was hidden from Swin.

The CNN and autoencoder have very similar MSE at validation. The encoder of the autoencoder and the structure of the CNN are identical so this result may not be surprising. The aim of using an autoencoder was to determine if the requirement to reconstruct the input image would help encode time information into the latent neuron being used to estimate the merger time. As the CNN and autoencoder results are so similar, this evidently was not achieved. This suggests that the CNN was extracting the most information from the images and the further image reconstruction did not provide further guidance.

While developing our CNN architecture, the performance of the CNN was seen to improve as more convolutional layers were added. With the methodology used in our CNN, adding a pooling layer after every convolutional layer, it is not possible to add further convolutional layers as the size of the image is already reduced to $2\time2$ pixels. It may be possible to find better results with more convolutional layers by not having pooling after every convolution.

As can be seen in Fig. \ref{fig:discuss:separation}, none of the neural network are predicting merger times, for pre-mergers, that are correlated with the projected separation between the merging galaxies. This suggests that the merger times being derived by all network are a result of morphological disturbances to the galaxies caused by their interactions with one-another.

\begin{figure}
	\resizebox{\hsize}{!}{\includegraphics{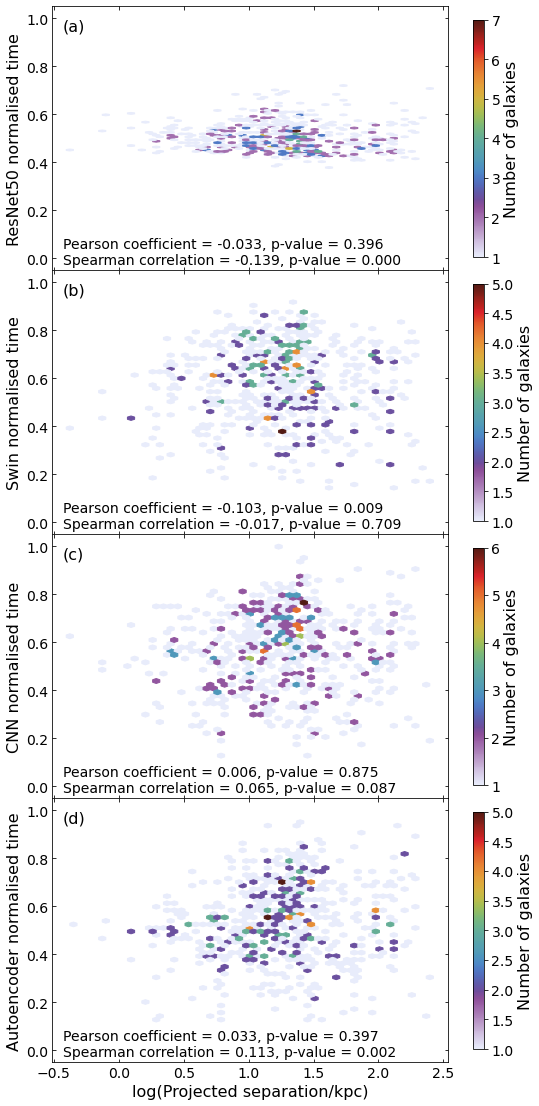}}
	\caption{Predicted merger times of the validation dataset from (a) ResNet50, (b) Swin, (c) CNN, and (d) autoencoder against projected separation of merging galaxies at the time of imaging for pre-mergers. Colour corresponds to number density from low (purple) to high (red).}
	\label{fig:discuss:separation}
\end{figure}

\subsubsection{Errors as a function of normalised time}\label{sec:discuss:nn-error}
The `U' shape of the mean and median errors as a function of normalised time for ResNet50 (Fig. \ref{fig:res:time-error}, dark blue downward pointing triangles) is a result of this architecture only outputting values close to 0.5 and is therefore not informative.

For the remaining three networks, the increases of the errors at low and high normalised merger time is also seen in Fig. \ref{fig:res:time-nn}, with the predictions at low true normalised times being over-estimated and the predictions and high true normalised times being under-estimated. As all three networks fail in the same way, it suggests that the reasons for failure are similar. At low normalised merger times ($< 0.33$), we have post-merger galaxies. As these galaxies relax after their merger, the evidence of a merger event becomes less obvious. Thus, all three networks begin to have difficulty discerning between mergers that happened more than approximately 125~Myr ago, indicated by the flat trend with true normalised times $\lesssim 0.25$ in Fig. \ref{fig:res:time-nn}b, c, and d.

For pre-mergers (normalised true time $> 0.33$), there is a similar problem. For the very early mergers, with true normalised time $\gtrsim 0.75$, Swin, CNN, and autoencoder again struggle to distinguish between the galaxies. This can again be seen in Fig. \ref{fig:res:time-nn}b, c, and d as a systematic underestimation at these high merger times. With all three networks failing similarly, this suggests that the influence of the merger events at these early times are not strong enough to be easily discernable. This would imply that galaxies between $\approx 625$ and $\approx 1000$~Myr before a merger event cannot be distinguished from one-another. However, as the majority of galaxies in Fig. \ref{fig:res:time-nn} at very high and very low merger times are not being confused for one-another, early pre-mergers and late post-mergers are distinguishable from each other. This is seen in Fig. \ref{fig:res:time-nn} as galaxies with very high true normalised merger times ($\approx 1$) are not being systematically predicted to have very low normalised merger times ($\approx 0$) and vice versa. This supports our choice to not enforce the companion galaxy of pre-mergers being within the image.

\subsubsection{Occlusion}\label{sec:discuss:occlusion}
To better understand which features the networks use to determine the time before or after the merger event, we perform an occlusion experiment. Here we use a $8\times8$ pixel square in the top left corner of the image and set all the values in the square to zero in all channels. The square of occluded pixels is transposed by 1 pixel in the x-direction or y-direction and the process repeated until the square reaches the bottom right corner of the image. This process produces 14\,641 versions of each galaxy. These images are then processed by the networks, creating a prediction of the merger time with the occluded pixels. 

Using these 14\,641 times, we can create a heat map. Each pixel in the heat map shows the average merger time of the galaxy from every instance where that pixel was occluded. We create these maps individually for all four networks.

We perform the occlusion experiment on five validation galaxies. These galaxies were selected to be the three pre-merger and two post-merger galaxies with the lowest average squared merger time error from Swin, CNN and autoencoder. We do not include ResNet50 due to it only providing output values close to 0.5. Using the occluded merger times, we generate Fig. \ref{fig:occlusion} which shows the original galaxy (left) followed by the ResNet50, Swin, CNN, and autoencoder occlusion heat-maps (left to right). Galaxies a, b, and c in Fig. \ref{fig:occlusion} are pre-mergers while d and e are post-mergers.

\begin{figure*}
	\resizebox{\hsize}{!}{\includegraphics{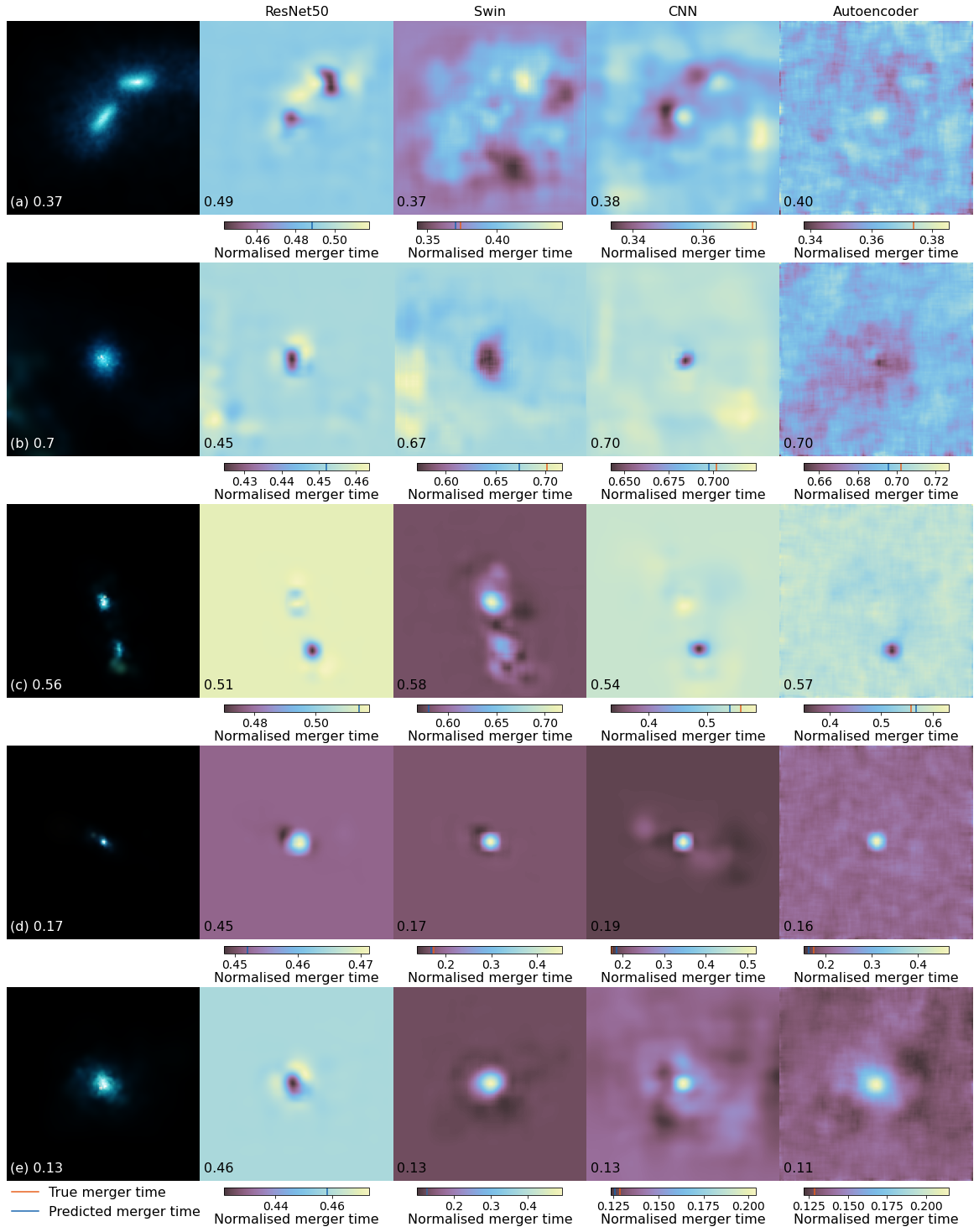}}
	\caption{Heat-maps with each pixel showing average merger time of when it is occluded from ResNet50, Swin, CNN, and autoencoder as labelled by column title. Each heat-map is colour scaled individually, where dark colours indicate an earlier time and light colours indicate a later time. True time is shown as red line in the colour bar while the un-occluded predicted time is shown as a yellow line. Original galaxy is shown on the left of each row as $u$, $g$, and $r$ band composite with each channel individually arcsinh scaled twice and then normalised between 0 and 1. The true normalised merger time is shown in the left panel of each galaxy while the predicted normalised time is shown in each networks' heat-map.}
	\label{fig:occlusion}
\end{figure*}

The ResNet50 occlusion does not overlap with the true time for any of the five galaxies. This further supports the poor performance of this network. What we do see is when the secondary galaxy is occluded, the normalised merger time is reduced with respect to when other regions are occluded. As later stage pre-mergers have two closely spaced galaxies, that may not be separable, and post mergers are single galaxies, having only one galaxy in the image would suggest a lower normalised merger time, which agrees with what is seen. We also see that the occlusion of faint structures around the secondary galaxies, in the case of a, b, and c, or the primary galaxy, in the case of a, b, c, and e, results in an increase in the normalised merger time with respect to when other regions are occluded. This makes intuitive sense for pre-mergers. A larger merger time indicates a longer period before the merger event and so the galaxies will be less disturbed by one-another. Hence, less disruption would imply more time until the merger for pre-mergers. For post-mergers, less disruption would mean more time since a merger and so occluding faint structures should decrease the normalised merger time, as seen with d. However all occluded normalised merger times are within $\approx0.06$ of each other, further demonstrating the problems with ResNet50.

Unlike ResNet50, Swin sees an increase in normalised merger time when the secondary galaxy is occluded with respect to when other regions are occluded. Swin does see an increase in merger time when faint structures are occluded for a, b, and c, with respect to when other regions are occluded, as would be expected for pre-mergers. Similarly, when the faint structures of d and e are occluded we see a decrease in normalised merger time with respect to when other regions are occluded, again as would be expected for post-mergers. For galaxies d and e, hiding the centre of the images, where the two nuclei of the merging galaxies may be present, increases the normalised merger time. Therefore, Swin is looking at nearby galaxies and faint structures to determine the time before a merger event.

The CNN was able to reproduce the correct and predicted merger times for all the galaxies studied here. Again, this network is using nearby galaxies and faint structures to determine the merger time. Hiding nearby galaxies reduces the normalised merger time with respect to when other regions are occluded while hiding fainter structures increases the normalised merger time for pre-merger galaxies. Hiding the faint structures around post-merger galaxies decreases the normalised merger time. Like with Swin, hiding the centres of galaxies d and e increases their merger time. Occluding the primary galaxy, for a and c, increases the merger time prediction while hiding the centre of b reduces the merger time. This suggests that there are features in the centre of interacting galaxies that give clues to how long there is until the galaxies will merge, for pre-mergers.

The final network, the autoencoder, appears to be picking up information from the background. This is surprising as the idealised images used contain no background noise. Again, the autoencoder finds information from the secondary galaxies but relied less on faint structures. As with Swin and CNN, the centre of galaxies d and e is important for finding a higher normalised merger time. Like with the CNN, hiding the secondary galaxy in c makes the predicted normalised merger time too low. Also like the CNN, occluding the primary galaxy in b decreases the estimate of the normalised merger time, again indicating that there is information in these regions that helps determine the time before a merger event.

\subsection{Latent space}\label{sec:discuss:latent}
With the autoencoder not performing better than the CNN, it may be possible that the latent space holds information about the time before, or after, a merger is that can be extracted. Evidently this is not the case as all six of the dimensionality reduction routines produce worse results than the neural networks, except ResNet50. However, this does not exclude the possibility that extra information is within the latent space, just that it is not a key component of the encoding into lower dimensions. This is despite LDA and NCA using the true merger times when creating the encoding.

No latent space encodings only produce normalised merger times close to 0.5, as can be seen in Fig. \ref{fig:res:time-latent}. With all MSE being worse than ResNet50, this suggests that ResNet50 is actively failing to find a mapping between the images and merger time. However, there is an indication that Isomap, LDA, SRP, and TSVD do provide similar output values at true normalised times above 0.4, although the scatter is too large in these regions to give a firm statement.

Examining the x-embedding of LDA provides a MSE close to that of the autoencoder, and CNN. This suggests that the x-embedding is a close reproduction of the time neuron in latent space. This is not too surprising as LDA uses the truth time to help encode the 64-dimension latent space into 2 dimensions. However, the slight decrease in MSE indicates that other dimensions in latent space are also being encoded into the x-encoding.

\subsubsection{Errors as a function of normalised time}\label{sec:discuss:latent-error}
The increase in mean and median errors for the predicted merger times for the latent space embeddings shows that these methods cannot distinguish between very late post-mergers and also between very early pre-mergers. This can be seen in Fig. \ref{fig:res:time-latent} with over estimation for low true normalised merger times and underestimation for high true normalised merger times. The late post-mergers are, however, do appear distinguishable from the early pre-mergers. Although the large scatter of NCA and TSVD make this difficult to be certain for these embeddings. This is unsurprising as the same is seen for the autoencoder, whose latent space encoding is used for the latent space embedding. With all the embeddings showing worse performance compared to the autoencoder, it is not surprising to see larger mean and median errors at all normalised merger times compared to the autoencoder.

The three worse performing embeddings, NCA, SRP, and TSVD, are also the three embeddings that have error evolution closest to a `U' shape. As discussed above and seen for ResNet50, the `U' shape can be caused by normalised merger time estimates all being close to a single value. However, Fig. \ref{fig:res:time-latent}c, d, and e do not show evidence of this, just a large scatter. Thus these three methods appear to be a poor choice for recovering the merger time.

The remaining three encodings, Isomap, LDA, and UMAP, show a peak in their errors at a normalised time of $\approx 0.35$. This bin contains the merger event (0.33). As can be seen in Fig. \ref{fig:res:time-latent}, these three embeddings seem to over estimate the merger time for galaxies with a true normalised merger time just above 0.35 with this not being so much of an issue for galaxies with a normalised merger time less than 0.35. This would indicate that the galaxies very close to merging are being assigned times earlier in the sequence than they should be, causing a peak in the errors seen in Fig. \ref{fig:res:latent-error}. This may be due to the presence of a second nucleus in these images of galaxies just about to complete merging fooling the embedding into providing a larger normalised time. The just completed mergers do not have a second nucleus and so do not suffer from this problem.

\subsection{Test data}\label{sec:discuss:test}
With a median error of 190~Myr, the CNN is able to produce a merger time that is approximately one snapshot away from the truth. The average difference in universe age between the snapshots used in this work is 162~Myr, not much smaller than the median error, and the largest age difference between the snapshots used is 194~Myr. This suggests that, at least for the data quality used in this work, it may not be beneficial to attempt to increase the time resolution through simple gravity simulations. Our requirements of the simple simulations between snapshots finding a merger time reduces our sample size by almost a factor of two: 3321 galaxies available after simulation compared to 6139 galaxies before. More training data may be beneficial in improving the median errors compared to a finer time resolution. However, tests with the CNN do not show an improved performance using a larger sample of galaxies without the simple gravity simulations applied.

The trends of the errors for the test data being similar to the validation data is expected, along with larger errors. This means that the galaxies furthest away from a merger (more than 125~Myr since a merger and longer than 625~Myr before a merger) are most likely to be miss-determined. As can be seen in Fig. \ref{fig:res:test-time}, this miss-determination is typically not early pre-mergers being given times consistent with late post-mergers or vice-versa, as was also seen with the validation data. As with the validation data, this means that the CNN struggles to find differences between very early mergers and differences between very late stage mergers. For the intermediate galaxies, with normalised merger times between 0.25 and 0.55, there is not a large loss of quality in the time predictions, when compared to the validation data. This reinforces the idea that the very early and very late stage mergers are the most difficult to reliably assign merger times. If we limit our true normalised merger times to be between 0.25 (125~Myr since a merger) and 0.75 (625~Myr before a merger), the CNN has a MSE of 0.022 for the validation data and 0.024 for the test data. These correspond to a mean error of 173~Myr and a median error of 144~Myr for the validation data and 182~Myr mean error and 157~Myr median error for the test data.

Our test result time error is not performing as well as \citet{2021arXiv210205182K}, who find a median error of 69.35~Myr. However, \citet{2021arXiv210205182K} use a much larger number of simulated galaxies, 203\,667 compared to our 9963, a smaller time range, -400 to 400~Myr compared to our -500 to 1000~Myr, and a higher angular resolution, 0.06 arcsec/pixel compared to our 0.20 arcsec/pixel. A larger amount of data provides more information to better train a neural network. Smaller time ranges make the regression problem simpler, as noted above the longer times before or after the merger event are more difficult to accurately predict merger times for. Finally, greater spatial resolution allows finer details to be detected by a neural network, with the faint structures being important for accurately determining the merger time as shown above. The data used by \citet{2021arXiv210205182K} also has a higher time resolution by a factor of 10, with a time resolution of $\approx 17$~Myr compared to our $\approx 162$~Myr before we perform the simple gravity extrapolations between snapshots. On a per-snapshot basis, we are performing better than \citet{2021arXiv210205182K}, with \citet{2021arXiv210205182K} achieving a median error of 4.0 snapshots compared to our median error of 1.2 snapshots. Our larger mis-determinations of very early stage pre-mergers is also seen in \citet{2021arXiv210205182K}.

Our CNN finds 81$\pm$1\% of our test data to be correctly classified as pre- or post-mergers. Again this is slightly lower than the 86\% accuracy found by \citet{2021arXiv210205182K} for the same binary classification task. However, our pre- and post-merger times are asymmetric, from -500~Myr to 1000~Myr, compared to the symmetric time range of -400~Myr to 400~Myr. Thus we have more pre-mergers than post-mergers which will influence our results. Indeed, 90$\pm$1\% of our pre-mergers are correctly identified while only 65$\pm$3\% of our post-mergers receive the correct classification. With our large uncertainties and smaller time range for a post-merger compared to a pre-merger, it is not surprising to see worse classification for post-mergers than pre-mergers.

\section{Conclusion}\label{sec:conc}
In this work, we explore how we can determine the time before or after a merger event a merging galaxy is. This was done by testing a residual network (ResNet50), a Swin Transformer (Swin), a convolutional neural network (CNN), and an autoencoder trained on images of merging galaxies from the IllustrisTNG TNG100. The merging galaxies were between 500~Myr after and 1000~Myr before a merger event. We also explore the latent space of the autoencoder.

The CNN was found to perform the best. We also find that the CNN and autoencoder have similar performance, suggesting that requiring the image of the galaxy to be reconstructed does not help encode merger time information. We find that ResNet50 performs the worse, possibly due to the relatively small sample size used for training. To add to this, the very early stage mergers, more than $\approx 625$~Myr before a merger event, and very late stage mergers, more than $\approx 125$~Myr since a merger, are difficult for any of the networks to accurately identify, possibly due to only very faint merger features present in these galaxies.

We also examined six different dimensionality reduction techniques: Isomap, linear discriminant analysis (LDA), neighbourhood components analysis (NCA), sparse random projection (SRP), truncated singular value decomposition (TSVD), and uniform manifold approximation and projection (UMAP). By mapping the validation data into the 2-dimensional latent space embedding proved to function poorer than the neural networks, with the exception of ResNet50. However, where ResNet50 produced normalised merger time predictions close to 0.5, the results from the latent space did not. We also found that the LDA x-embedding was a good approximation of the autoencoder latent space neuron used to estimate the merger time. By converting the x-embedding into a merger time, the results were similar to the autoencoder itself.

The best performing method, the CNN, produced merger times with a median error of 190~Myr. This time is similar to the time between snapshots in the IllustrisTNG simulation. The errors may be improved by reducing the time range we consider to be mergers to 125~Myr after to 625~Myr before the merger event.

While the methods in this work did not produce a highly accurate reproduction of the merger time, they show that is is possible to derive this information from idealised images of galaxy mergers. Future work will look at how our results can be improved. This will be done by exploring how a larger number of mergers from simulations may help improve performance as well as reducing the time range that can be considered to be a merger. It will also look at if increasing the spatial resolution of the images, the depth of the images, and time resolution of the simulation will aid with merger time determination. Future work will also explore if more realistic images of simulated galaxies will result in poorer performance and if including morphological parameters can improve performance. This future work will aim to derive the merger time for observed galaxy mergers, allowing changes in physical properties to be statistically tracked throughout a merger event. As the CNN is the least computationally intensive network to train, and its performance compared to the other networks in this work, these future works will be based on our CNN architecture. 

\begin{acknowledgements}
	We would like to thank the referee for their thorough and thoughtful comments that helped improve the quality and clarity of this paper.
	
	W.J.P. has been supported by the Polish National Science Center project UMO-2020/37/B/ST9/00466.
	
	The IllustrisTNG simulations were undertaken with compute time awarded by the Gauss Centre for Supercomputing (GCS) under GCS Large-Scale Projects GCS-ILLU and GCS-DWAR on the GCS share of the supercomputer Hazel Hen at the High Performance Computing Center Stuttgart (HLRS), as well as on the machines of the Max Planck Computing and Data Facility (MPCDF) in Garching, Germany.
\end{acknowledgements}

\bibliographystyle{aa} 
\bibliography{OPUS-sim} 

\begin{appendix}
\section{Latent space embeddings}\label{app:embeddings}
Here we present the latent space embeddings for Isomap, NCA, SRP, TSVD, and UMAP in Figs. \ref{fig:res:isomap}, \ref{fig:res:nca}, \ref{fig:res:srp}, \ref{fig:res:tsvd}, and \ref{fig:res:umap}, respectively, using the latent space from the autoencoder epoch with the lowest merger time MSE. The LDA embedding can be found in Fig. \ref{fig:res:lda}.

\begin{figure}
	\resizebox{\hsize}{!}{\includegraphics{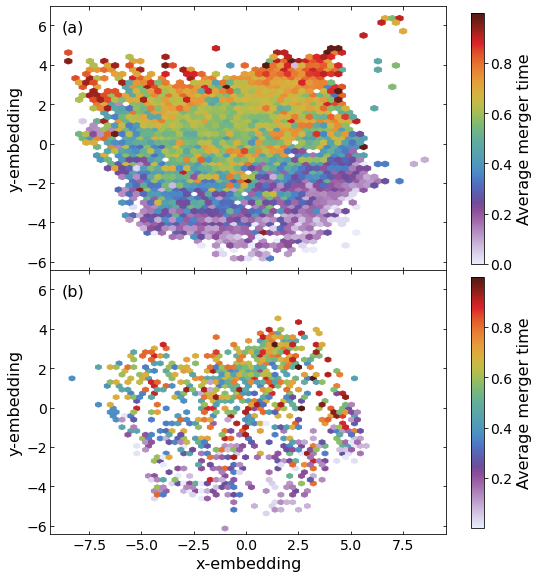}}
	\caption{Isomap latent space embedding for (a) training data and (b) validation data. The average normalised merger time is show from 0 (purple) to 1 (red).}
	\label{fig:res:isomap}
\end{figure}

\begin{figure}
	\resizebox{\hsize}{!}{\includegraphics{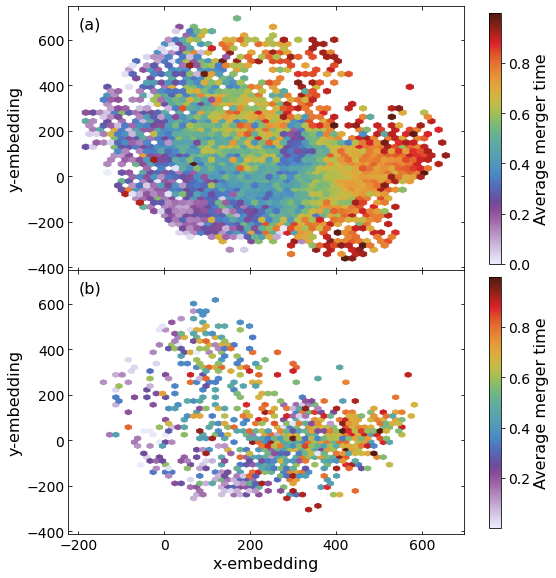}}
	\caption{NCA latent space embedding for (a) training data and (b) validation data. The average normalised merger time is show from 0 (purple) to 1 (red).}
	\label{fig:res:nca}
\end{figure}

\begin{figure}
	\resizebox{\hsize}{!}{\includegraphics{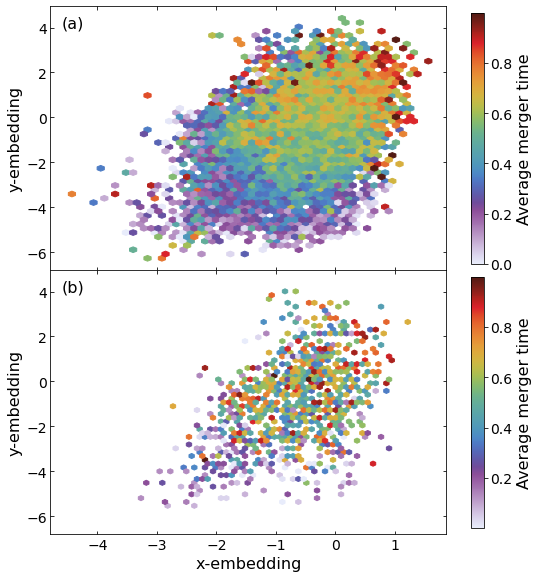}}
	\caption{SRP latent space embedding for (a) training data and (b) validation data. The average normalised merger time is show from 0 (purple) to 1 (red).}
	\label{fig:res:srp}
\end{figure}

\begin{figure}
	\resizebox{\hsize}{!}{\includegraphics{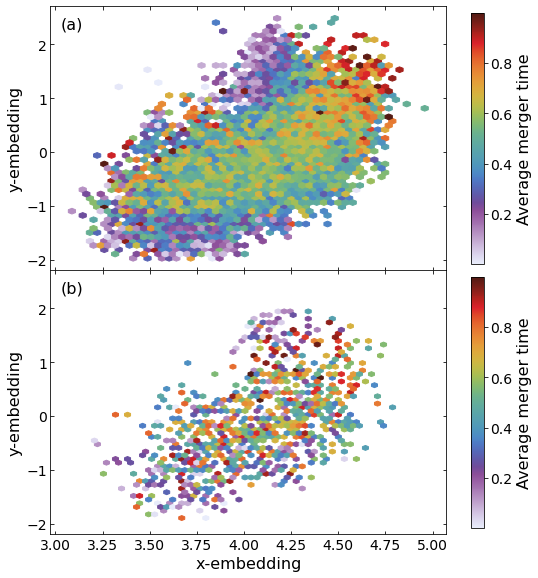}}
	\caption{TSVD latent space embedding for (a) training data and (b) validation data. The average normalised merger time is show from 0 (purple) to 1 (red).}
	\label{fig:res:tsvd}
\end{figure}

\begin{figure}
	\resizebox{\hsize}{!}{\includegraphics{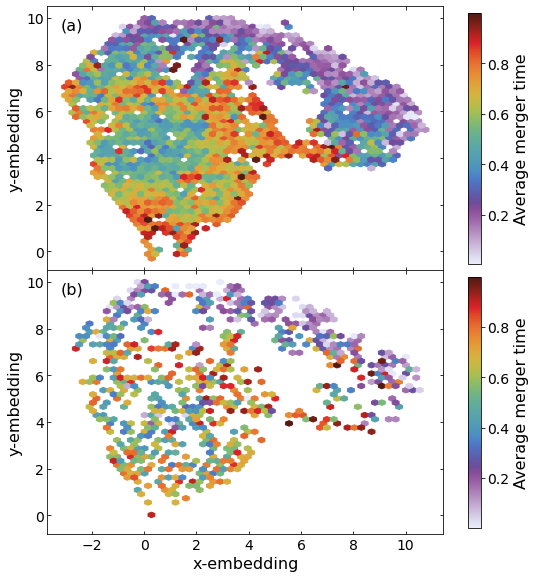}}
	\caption{UMAP latent space embedding for (a) training data and (b) validation data. The average normalised merger time is show from 0 (purple) to 1 (red).}
	\label{fig:res:umap}
\end{figure}

\end{appendix}

\end{document}